\documentclass[10pt,]{asme2ej}

\usepackage{graphicx}
\usepackage{amsmath, amssymb, latexsym}

\title{Discrete Filters for Large Eddy Simulation of Forced Compressible MHD Turbulence}

\author{Alexander A. Chernyshov\\
    \affiliation{
	Space Research Institute \\of Russian
Academy of Sciences\\
Profsoyuznaya 84/32, 117997\\ Moscow, Russia\\
    Email: achernyshov@iki.rssi.ru
    }	
}

\author{Kirill. V. Karelsky \\
    \affiliation{   Space Research Institute \\of Russian
   Academy of Sciences\\
     Profsoyuznaya 84/32, 117997\\ Moscow, Russia\\
     Email: kkarelsk@iki.rssi.ru
    }
}

\author{Arakel. S. Petrosyan \thanks{Address all correspondence to this author.}\\
           \affiliation{
    Space Research Institute \\of Russian
   Academy of Sciences\\
     Profsoyuznaya 84/32, 117997\\ Moscow, Russia\\
           }
        \affiliation{
    Moscow Institute of Physics and Technology \\
    State University\\
   Institutskiy Pereulok 9, 141700\\ Moscow Region, Dolgoprudny, Russia\\
    Email: apetrosy@iki.rssi.ru\\
    }
    }

\begin{document}

\maketitle

\begin{abstract}
{\it
In present study, we discuss results of applicability of discrete filters for large eddy simulation (LES) method of forced compressible magnetohydrodynamic (MHD) turbulent flows with the scale-similarity model. Influences and effects of discrete filter shapes on the scale-similarity model are examined in physical space using a finite-difference numerical schemes. We restrict ourselves to the Gaussian filter and the top-hat filter. Representations of this subgrid-scale model which correspond to various 3- and 5-point approximations of both Gaussian and top-hat filters for different values of parameter $\epsilon$ (the ratio of the mesh size to the cut-off lengthscale of the filter) are investigated. Discrete filters produce more discrepancies for magnetic field. It is shown that the Gaussian filter is more sensitive to the parameter $\epsilon$  than the top-hat filter in compressible forced MHD turbulence. The 3-point filters at $\epsilon=2$ and $\epsilon=3$ give the least accurate results and the 5-point Gaussian filter shows the best results at $\epsilon=2$.

}
\end{abstract}

%

\section{Introduction}

Compressible turbulent flows in a
magnetic fields are common both in engineering and applied areas and in physics of
astrophysical and space processes. Among the engineering
applications, possibility of boundary layer control and drug
reduction, MHD flow in channel for steel-casting process, and in pipe for cooling
of nuclear fusion reactors can be mentioned. Most of the
applications demand understanding of turbulent flow at high
Reynolds numbers with density fluctuations due to compressibility
(like for example, in aerospace engineering design). The presence of velocity and magnetic field fluctuations in a wide range of space and time scales has been directly detected in the various turbulent flows in space processes. For example, there are strong indications of their presence in the solar corona, interplanetary medium, solar wind and others. Note that the MHD problems differ from those of the neutral fluid hydrodynamics. The MHD equations contain
two fields which introduces considerably more freedom into the dynamics. Fundamental limitations of direct numerical simulation
(DNS) method for a turbulence modeling and difficulties due to
presence of compressibility and magnetic field demand development
of new theoretical and computational methods and make important
advancing of large eddy simulation (LES) method for such complex
MHD flows. According to LES approach, the large-scale part of the
flow is computed directly and only small-scale structures of
turbulence are modeled. This scale separation is achieved by applying
a filter. The small-scale motion is eliminated from
the initial system of equations of motion by filtering procedures
and its effect is taken into account by special closures referred
to as the subgrid-scale (SGS) models \cite{GarAdamsSag09,Biskamp2003,Cher06POP,Cher08POF,Cher07POF,Car02,Cher09TCFD}.

Theoretical studies on SGS modeling are carried out by performing filtering
operations that are defined as convolution products between the velocity field and the filter
kernel. Such definition is suitable when dealing with numerical methods such as spectral or
pseudo-spectral types, and is expensive when dealing with local methods (finite
differences, finite volumes, finite elements). In practice, for local methods, discrete test filters
with compact stencils based on weighted averages are used.
The properties of the discrete filters differ substantially from those of the continuous filters, which
are the basis of theoretical analysis. Hence, the need for the analysis of discrete filters, and for
the choice of discrete filters with required properties in order to ensure a greater consistency
between the continuous SGS model and its discretized version, which will be used for the
computations.

In the present paper, we deal with the question of the effects and influences of different filter shapes on scale-similarity model in LES method for compressible forced MHD turbulent flows using a finite-difference schemes. Recently, we have showed that the scale-similarity model for forced MHD turbulence can be used as a stand alone SGS model as opposed to decaying case \cite{Cher12FTC}. The scale-similarity parametrization has evident advantages the main ones being to reproduce rightly the correlation between the tensors between actual and model turbulent stress tensor for isotropic flow as well as for anisotropic fluid flow, and the absence of special model constants in contrast to other SGS closures. However, the scale-similarity model does not dissipate energy enough and usually leads to inaccurate results in decaying turbulence or blows up the simulation. But the situation changes significantly when a forced turbulence is considered. In this case, subgrid modeling in LES approach provides correct stationary regime of the turbulence rather than to guarantee proper energy dissipation. It was shown that the scale-similarity model provides good accuracy and the results of this SGS model agree well with the DNS results. If differences between the results obtained by the scale-similarity model and the Smagorinsky closure for velocity field are insignificant, then the differences are considerable for magnetic field. For the magnetic field, discrepancies with the DNS results are substantially lower for scale-similarity model while the Smagorinsky parametrization for MHD case is more dissipative and the results of Smagorinsky model are worse in agreement with DNS\cite{Cher12FTC}. The scale-similarity model is generally found to reproduce DNS results better.

The present paper briefly summarizes results concerning discrete filters for LES of forced compressible MHD turbulence by the example of scale-similarity model. The structure of the paper is the following. The next section 2 describes the general features of LES technique in physical space. Influence and sensitivity of discrete filter shapes on scale-similarity model, test configurations and numerical analysis of the obtained results are specified in section 3. Finally, conclusion remarks are given in the last section 4.

\section{Filtering procedure in large eddy simulation}

In this section, we formulate the general features of the theory of LES method for modeling of
compressible forced MHD turbulent flows.

To obtain the MHD equations governing the motion of the filtered
(that is resolved) eddies, the large scales from the small are
separated. LES approach is based on the definition of a filtering
operation: a resolved (or large-scale) variable, denoted by an
overbar in the present paper, is defined as

\begin{equation}\label{E:filtr_def}
  \bar{\zeta}(x_{i})=\int_{\Theta} \zeta(\acute{x_{i}}) \xi (x_{i}, \acute{x_{i}};
  \bar{\triangle}) d\acute{x_{i}},
\end{equation}
where $\xi $ is the filter function satisfying the normalization
property, $\zeta$ is a flow parameter, $\Theta$ is the domain, $\bar{\triangle}$ is the
filter-width associated with the wavelength of the smallest scale
retained by the filtering procedure and $x_{j} = (x,y,z)$ are axes
of Cartesian coordinate system.

It is convenient to use the Favre filtration (it is also called
mass-weighted filtration) to avoid additional SGS terms in
compressible flow. Therefore, Favre filtering will be used further. Mass-weighted filtering is used for all parameters of charged fluid
flow besides the pressure and magnetic fields. Favre filtering is determined as follows:
\begin{equation}
   \tilde{\zeta} = \frac{\overline{{\rho} \zeta}} {\bar{\rho}} \label{E:favre}
\end{equation}
where the tilde denotes the mass-weighted filtration.

Thus, applying the Favre-filtering operation, we can rewrite the MHD
equations for compressible fluid flow as \cite{Cher10POP,Cher12FTC}:

\begin{equation}\label{E:fneraz1}
  \frac{\partial{\bar{\rho}}}{\partial t} + \frac{\partial{\bar{\rho}} \tilde{u_{j}}}{\partial x_{j}} =
  0;
\end{equation}
\begin{equation}\label{E:fNS1}
  \frac{\partial{\bar{\rho}} \tilde{u_{i}}}{\partial t} + \frac{\partial}{\partial
  x_{j}}\left (\bar{\rho} \tilde{u_{i}} \tilde{u_{j}} + \frac{\bar{\rho}^{\gamma}}{\gamma M_{s}^{2}} \delta_{ij}
  -\frac{1}{Re}~\tilde{\sigma_{ij}} + \frac{\bar{B^{2}}}{2M_{a}^{2}}\delta_{ij} - \frac
{1}{M^{2}_{a}}~\bar{B_{j}} \bar{B_{i}}
  \right) =
  - \frac {\partial \tau_{ji}^{u}}{\partial x_{j}} + \tilde{F_{i}^{u}};
\end{equation}
\begin{equation}\label{E:fin1}
  \frac{\partial \bar{B_{i}}}{\partial t} + \frac {\partial}{\partial
  x_{j}}\left ( \tilde{u_{j}} \bar{B_{i}} -
  \tilde{u_{i}}\bar{B_{j}} \right )- \frac{1}{Re_{m}} \frac{\partial^{2} \bar{B}_{i}}{\partial x_{j}^{2}} =
   - \frac {\partial \tau_{ji}^{b}}{\partial
  x_{j}} + \tilde{F_{i}^{b}};
\end{equation}
\begin{equation}\label{E:divB1}
\frac{\partial \bar{B_{j}}} { \partial x_{j}} = 0,
\end{equation}
Here $\rho$ is the density;  $u_{j}$ is the velocity in the
direction $x_{j}$; $B_{j}$ is the magnetic field in the direction
$x_{j}$; $\sigma_{ij} = 2\mu S_{ij} - \frac{2}{3}\mu
S_{kk}\delta_{ij}$ is the viscous stress tensor; $S_{ij} =
1/2\left(
\partial u_{i} / \partial x_{j} + \partial u_{j}/ \partial x_{i}
\right)$ is the strain rate tensor; $\mu$ is the coefficient of
molecular viscosity; $\eta$ is the coefficient of magnetic
diffusivity; $\delta_{ij}$ is the Kronecker delta.

The filtered magnetohydrodynamic equations
~(\ref{E:fneraz1})~-~(\ref{E:divB1}) are written in the
dimensionless form using the standard procedure \cite{Biskamp2003}
where $Re= \rho_{0}u_{0}L_{0} / \mu_{0}$ is the Reynolds number,
$Re_{m} = u_{0}L_{0} / \eta_{0}$ is the magnetic Reynolds number.
$ M_{s}= u_{0} / c_{s}$ is the Mach number, where $c_{s}$ is the
velocity of sound defined by the relation $c_{s}= \sqrt{\gamma
p_{0} / \rho_{0}}$, and $M_{a} = u_{0} / u_{a}$ is the magnetic
Mach number, where $u_{a}= B_{0}/ (\sqrt{4\pi\rho_{0}})$ is the
Alfv{\'e}n velocity. To close the MHD equations ~(\ref{E:fneraz1})~-~(\ref{E:fin1}) it is
assumed that the relation between density and pressure is
polytropic and has the following
form: $p = \rho^{\gamma}$, where $\gamma$ is a polytropic index.

The effect of the subgrid scales appears on the right-hand side of the governing MHD equations ~(\ref{E:fNS1})~-~(\ref{E:fin1}) through the SGS
stresses:

\begin{equation}
\tau_{ij}^{u} = \bar{\rho} \left ( \widetilde{u_{i} u_{j}} -
\tilde{u_{i}}\tilde{u_{j}}\right)- \frac {1}{M_{a}^{2}}\left(
\overline{B_{i} B_{j}} - \bar{B_{i}}\bar{B_{j}}\right) ;
\end{equation}
\begin{equation}
\tau_{ij}^{b} =  \left ( \overline{u_{i} B_{j}} -
\tilde{u_{i}}\bar{B_{j}}\right)- \left( \overline{B_{i} u_{j}} -
\bar{B_{i}}\tilde{u_{j}}\right).
\end{equation}

Thus, the filtered system of magnetohydrodynamic equations
contains the unknown turbulent tensors: $\tau_{ij}^{u}$ and
$\tau_{ij}^{b}$. To determine their components special turbulent
closures (models, parameterizations) based on large-scale values
describing turbulent magnetohydrodynamic flow must be used. The
main idea of any SGS closures used in LES is to reproduce the
effects of the subgrid scale dynamics on the large-scale energy
distribution, at that as a matter of fact Richardson turbulent
cascade is simulated. In order to close the system of MHD
equations, one should find such parameterizations for
$\tau_{ij}^{u}$ and $\tau_{ij}^{b}$ that would relate these
tensors to the known large-scale values of the flow parameters.

There are external driving forces $F^{u}_{i}$ and $F^{b}_{i}$ on the right-hand sides of equations~(\ref{E:fNS1})~-~(\ref{E:fin1}) respectively. Driving forces $F^{u}_{i}$ and $F^{b}_{i}$ which sustain turbulence are necessary to study statistically stationary flow and provide a stationary picture of the energy cascade and more statistical sampling. If energy is not injected into a turbulent flow, after some time this flow becomes laminar because of viscosity and diffusion. To sustain a three-dimensional turbulence, a driving force is employed to inject energy in the turbulent system to replace the energy which is dissipated on small spatial scales.

Recently, "linear forcing" was suggested and used for modeling of compressible MHD turbulence \cite{Cher10POP} with driving force in physical space. The idea essentially consists of adding a force proportional to the fluctuating velocity\cite{Lun03,Cher10POP,RosMen05,DiStefVas10}. Linear forcing resembles a turbulence when forced with a mean velocity gradient, that is, a shear. This force appears as a term in the equation for fluctuating velocity that corresponds to a production term in the equation of turbulent kinetic energy. In compressible MHD turbulence, system of MHD equations includes also the magnetic induction equation and in this case the driving force is proportional to the magnetic field in the magnetic induction equation \cite{Cher10POP}. Thus, linear external force can be interpreted as the production of magnetic energy due to the interaction between the magnetic field and the mean fluid shear.

The determination of the driving forces $F^{u}_{i}$ and $F_{i}^{b}$ in the momentum conservation equation and in the magnetic induction equation, respectively, are:

\begin{equation}\label{E:force_u}
F^{u}_{i} = \Theta \rho u_{i}
\end{equation}
 \begin{equation}\label{E:force_b}
F^{b}_{i} = \Psi B_{i}
\end{equation}
where $\Theta$ in (\ref{E:force_u}) is the coefficient which is determined from a balance
of kinetic energy for a statistically stationary state. The forcing function $F^{u}_{i} = \Theta \rho u_{i}$ in the physical space is equivalent to force all the Fourier modes in the spectral space. This is in fact the only difference from the standard spectral forcing when energy is added in to system only in the range of small wave numbers (wavenumber shell), that is, in integrated (large) scale of turbulence. The coefficient $\Psi$ in the expression (\ref{E:force_b}) is determined from the balance of magnetic energy for the statistically stationary state as well. More detailed derivation and information about linear forcing method in physical space for compressible MHD turbulent flows can be found in our article \cite{Cher10POP}.

The scale-similarity model as a subgrid-scale closure for compressible MHD case is of the form\cite{Cher07POF}:

\begin{equation}\label{E:ssm_hydro}
\tau_{ij}^{u} =
\bar{\rho}\left(\widetilde{\tilde{u_{i}}\tilde{u_{j}}}
   -\tilde{\tilde{u_{i}}}\tilde{\tilde{u_{j}}} \right) - \frac{1}{M_{a}^{2}}\left(\overline{\bar{B_{i}}\bar{B_{j}}}
   -\bar{\bar{B_{i}}}\bar{\bar{B_{j}}} \right)
\end{equation}
\begin{equation}\label{E:ssm_mag}
\tau_{ij}^{b} = \left(\overline{\tilde{u_{i}}\bar{B_{j}}}
   -\tilde{\tilde{u_{i}}}\bar{\bar{B_{j}}} \right) - \left(\overline{\bar{B_{i}}\tilde{u_{j}}}
   -\bar{\bar{B_{i}}}\tilde{\tilde{u_{j}}} \right)
\end{equation}

The scale-similarity model for MHD turbulence ~(\ref{E:ssm_hydro})~-~(\ref{E:ssm_mag})
can be calculated in a LES technique by means of the filtered variables in
contrast to eddy-viscosity parameterizations. Model constants in ~(\ref{E:ssm_hydro}) and (\ref{E:ssm_mag}) are not introduced as this would destroy the Galilean invariance of the expression\cite{Spez85}.
\section{Numerical analysis of sensitivity of scale-similarity model on the filter shape}

\begin{figure}[t]
\centerline {
\includegraphics [width=120mm]{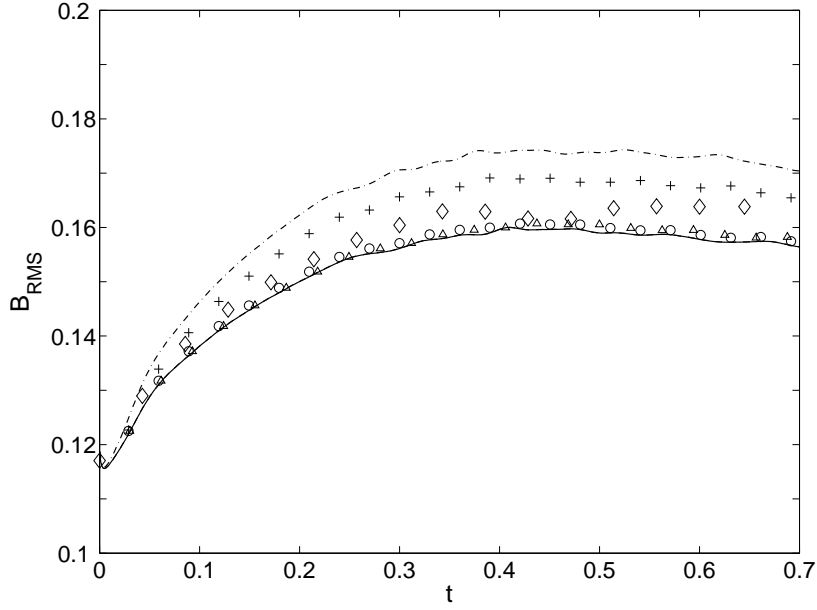}}
\caption {Time dynamics of $b_{rms}$ for various filter shapes. The diamond line is the DNS results, the solid line is 5-point approximation of the Gaussian filter ($\epsilon = 2$),
the dashed line is 5-point approximation of the top-hat filter ($\epsilon = 2$),
the dash-dot line is 3-point approximation of the Gaussian (or top-hat) filter ($\epsilon = 2$),
the circle line is 5-point approximation of the Gaussian filter ($\epsilon = 3$),
the triangle line is 5-point approximation of the top-hat filter ($\epsilon = 3$),
and the plus line is 3-point approximation of the Gaussian (or top-hat) filter ($\epsilon = 3$).}%
\label{Fi:b_rms_filtr}
\end{figure}

In this section, influences and sensitivity of discrete filter shapes on scale-similarity model, test configurations and numerical analysis of obtained results are studied. The  results obtained for LES are compared with the DNS results for three-dimensional forced compressible  MHD turbulent flows.

There is strong influence of the properties of the LES filter on the interactions between
resolved and subgrid-scales. We examine the question of the effect of different filter shapes on scale-similarity model for forced  compressible MHD turbulent flow using a finite-difference schemes. Several papers were devoted to this problem for a neutral fluids dynamics. Both theoretical and numerical studies have been carried out \cite{PiomFerMoin88,Sagaut99,LeslieQ79}. To our knowledge,  the influence of discrete filters on the scale-similarity model for the case of compressible forced MHD turbulence is not studied.

It should be remarked that the definition of filtering procedure (\ref{E:filtr_def}) is too general.
The real flows in the nature and in the experiments can be investigated with the help of some simpler appropriate filter. The operator defined by relation (\ref{E:filtr_def}) is a
priori non-local in physical space, and is then worst suited for computations performed with
local numerical methods (e.g. finite differences, finite elements, finite volumes). Therefore, it is necessary
to define some local discrete approximations for this operator. Since the finite-difference schemes for simulation of MHD turbulent flows are used in this paper, we consider the Gaussian filter and the top-hat (or box) filter. They are commonly applied when using non-spectral modeling techniques in physical space.

The top-hat filter is defined as:

\begin{equation}
\xi (x, \acute{x})= \begin{cases}
\frac{1}{ \bar{\triangle} }, \text{ if $\mid x - \acute{x}\mid \leq \frac{ \bar{\triangle} }{2}$ } \\
0, \text{otherwise}
\end{cases}
\end{equation}

The Gaussian filter is:

\begin{equation}
\xi (x, \acute{x}) = (\frac{6}{\pi \bar{\triangle}^{2}})^{1/2} exp (-\frac{6 \mid x - \acute{x}\mid ^{2}} {\bar{\triangle}^{2}})^{1/2} )
\end{equation}

Filter approach for hydrodynamics of neutral gas was analyzed by Sagaut and Grohens \cite{Sagaut99}. They were looking for an optimal shape of the filters which is consistent with the numerical scheme in use. They found by means of the Taylor series decomposition that the top-hat and Gaussian filters coincide exactly for second order accuracy numerical schemes (using 3 points):

\begin{equation}\label{E:filtr_3p}
\bar{\zeta_{i}} = \frac{1}{24}*\epsilon^{2}*\zeta_{i-1} + \frac{1}{12}*(12-\epsilon^{2})*\zeta_{i} +
\frac{1}{24}*\epsilon^{2}*\zeta_{i+1}
\end{equation}

Fourth order accuracy numerical schemes (using 5 point) consistent with different forms of these filters. Operator equivalent to the fourth-order Gaussian filter and top-hat filter respectively are:

\begin{equation}\label{E:filtr_g5}
\bar{\zeta_{i}} = \frac{\epsilon^{4} - 4\epsilon^{2}}{1152}(\zeta_{i-2} + \zeta_{i+2})  +%
\frac{16\epsilon^{2} - \epsilon^{4}}{288}(\zeta_{i-1} + \zeta_{i+1})%
+ \frac{\epsilon^{4} - 20\epsilon^{2} + 192}{192}\zeta_{i},
\end{equation}

\begin{equation}\label{E:filtr_b5}
\bar{\zeta_{i}} = \frac{3 \epsilon^{4} - 20 \epsilon^{2}}{5760}(\zeta_{i-2} + \zeta_{i+2})  +%
\frac{80\epsilon^{2} - 3 \epsilon^{4}}{1440}(\zeta_{i-1} + \zeta_{i+1})%
+ \frac{3\epsilon^{4} - 100\epsilon^{2} + 960}{690}\zeta_{i},
\end{equation}

Here, $\zeta_{i}$ is the flow parameter in the point $i$ and the
parameter $\epsilon$ represents the ratio of the mesh size to the
cut-off lengthscale of the filter \cite{Sagaut99}. It is usually assumed that the parameter $\epsilon$  is equal to $2$ in the works where the fluid flows are modeled by means of LES approach. However, in order to study how this parameter affects the results of the calculation, we consider the cases when the parameter $\epsilon$ takes a different value, namely, $\epsilon = 3$.

Initially it should be noted that since the problem considered in this work is three-dimensional three
dimensional filter (multidimensional one in the general case) must
be constructed. Multidimensional filter can be constructed in two
different ways \cite{Sagaut99}. The first one is a linear
combination of one-dimensional filters, i.e. for every direction the
flow parameter is filtered independently from the others

\begin{equation}\label{E:lin_fil}
\xi^{n} = \frac{1}{n} \sum_{i=1}^{n} \xi^{i},
\end{equation}

where $\xi^{i}$ is a one-dimensional filter in direction $i$, $n$ is
the number of space dimensions. Linear combination represents
simultaneous application of all one-dimensional filters in every
spatial direction. The second approach is a product of
one-dimensional filters. In that case the following can be written:

\begin{equation}\label{E:com_fil}
\xi^{n} =  \prod_{i=1}^{n} \xi^{i}.
\end{equation}

Such technique of determination of multidimensional filter $\xi^{n}$
represents non-simultaneous application of one-dimensional filters
like in the first case but sequential one. The accuracy of
constructed the multidimensional filters was tested by Sagaut and
Grohens \cite{Sagaut99}. They showed that sequential product of
filters gives more accurate results in comparison with linear
combination of one-dimensional filters. Therefore, in this work the
sequential product of filters ~(\ref{E:com_fil}) is used for
three-dimensional filtration.

\begin{figure}[t]
\centerline {
\includegraphics [width=120mm]{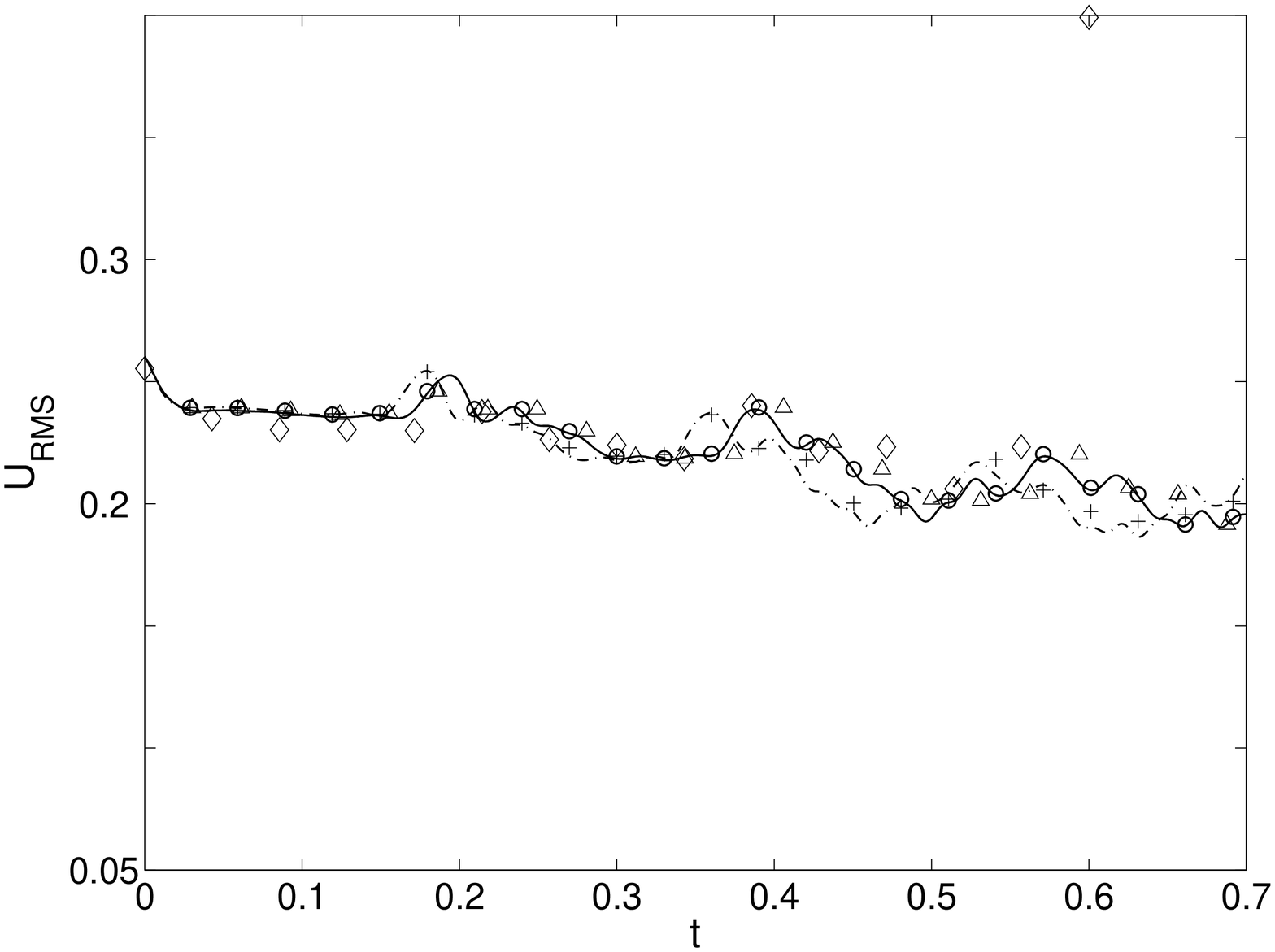}}
\caption {Time dynamics of $u_{rms}$ for various filter shapes. Symbols as in Fig.\ref{Fi:b_rms_filtr}}%
\label{Fi:u_rms_filtr}
\end{figure}

Three-dimensional numerical simulations of forced compressible MHD turbulence in physical space are performed and the numerical code of the fourth order accuracy for
MHD equations in the conservative form  based on non-spectral finite-difference
schemes  is used in our work. The third order low-storage Runge-Kutta method is
applied for time integration. The skew-symmetric form of nonlinear terms for modeling of turbulent flow is applied to reduce discretization errors. The skew-symmetric form is a form obtained by averaging divergent and convective forms of the nonlinear terms:

\begin{equation}\label{E:skew}
\Psi_{i}^{s} = \frac{1}{2} \left( \frac{(\partial \rho u_{i}u_{j})}{\partial x_{j}}  + \rho u_{j} \frac{ \partial u_{i}}{\partial
x_{j}} + u_{i}\frac{(\partial \rho u_{j})}{\partial x_{j}}
\right)
\end{equation}

In spite of analytical equivalence of all three forms, their numerical realizations give different results and it was shown that skew-symmetric form improves computational accuracy for turbulent modeling. Periodic boundary conditions for all the three dimensions are
applied. The similarity numbers in all simulations are: $Re \approx 300$, $Re_{M} \approx 50 $, $M_{s} \approx 0.35$, $M_{a} \approx 1.4$, $\gamma = 1.5 $. The simulation domain is a cube $\pi \times \pi \times \pi$. The mesh with
$64^{3}$ grid cells is used for LES and $256^{3}$ for DNS. The explicit LES method is used in this work. The initial isotropic turbulent spectrum close to $k^{-2}$ with random amplitudes and phases in all three directions was chosen for kinetic
and magnetic energies in Fourier space. The
choice of such spectrum as initial conditions is due to velocity
perturbations with an initial power spectrum in Fourier space
similar to that of developed turbulence \cite{Mac_Low98PRL}. This $k^{-2}$ spectrum corresponds to spectrum of Burgers turbulence. Initial conditions for the velocity and the magnetic field have been obtained in the physical space using inverse Fourier transform. The results obtained with LES technique are compared with
DNS computations and performance of large eddy simulation is
examined by difference between LES- and filtered DNS-results. The
initial conditions for LES are obtained by filtering the initial
conditions of DNS.

Since our interest is on study scale-similarity SGS models which rely on the application of a filter to its discrete formulation, we consider various versions of scale-similarity closure that correspond to various 3- and 5-point approximations of both Gaussian and top-hat filters for $\epsilon = 2$ and $\epsilon = 3$.

Time dynamics of root-mean-square magnetic field $b_{rms}$ and root-mean-square velocity $u_{rms}$ are shown in Fig.\ref{Fi:u_rms_filtr} and in Fig.\ref{Fi:b_rms_filtr} respectively. Here and below, in Fig.\ref{Fi:u_rms_filtr} and Fig.\ref{Fi:b_rms_filtr}, the diamond line is the DNS results,
the solid line is 5-point approximation of the Gaussian filter ($\epsilon = 2$),
the dashed line is 5-point approximation of the top-hat filter ($\epsilon = 2$),
the dash-dot line is 3-point approximation of the Gaussian (or top-hat) filter ($\epsilon = 2$),
the circle line is 5-point approximation of the Gaussian filter ($\epsilon = 3$),
the triangle line is 5-point approximation of the top-hat filter ($\epsilon = 3$),
and the plus line is 3-point approximation of the Gaussian (or top-hat) filter ($\epsilon = 3$).
In these plots, we can see that the use of the 5-point filters lead to a increase of the
accuracy. The largest discrepancy with the DNS results is observed for scale-similarity results with the 3-point Gaussian (or top-hat) filters at different values of $\epsilon$. At the same time, the 5-point filters are in good agreement with the "exact" results of DNS. One can notice that 5-point filters lead to similar results for the two
values of parameter $\epsilon$ whereas a 3-point filter produces more discrepancies for magnetic field.

\begin{figure}[t]
\centerline {
\includegraphics [width=120mm]{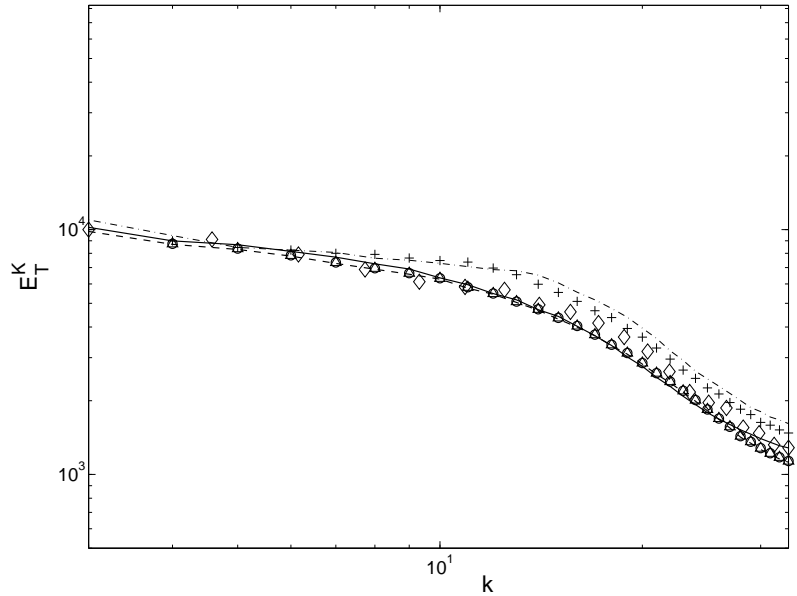}}
\caption {Total energy spectrum $E_{T}^{K}$ for various filter shapes. Symbols as in Fig.\ref{Fi:b_rms_filtr}}%
\label{Fi:tot_spect_filtr}
\end{figure}

\begin{figure}[t]
\centerline {
\includegraphics [width=120mm]{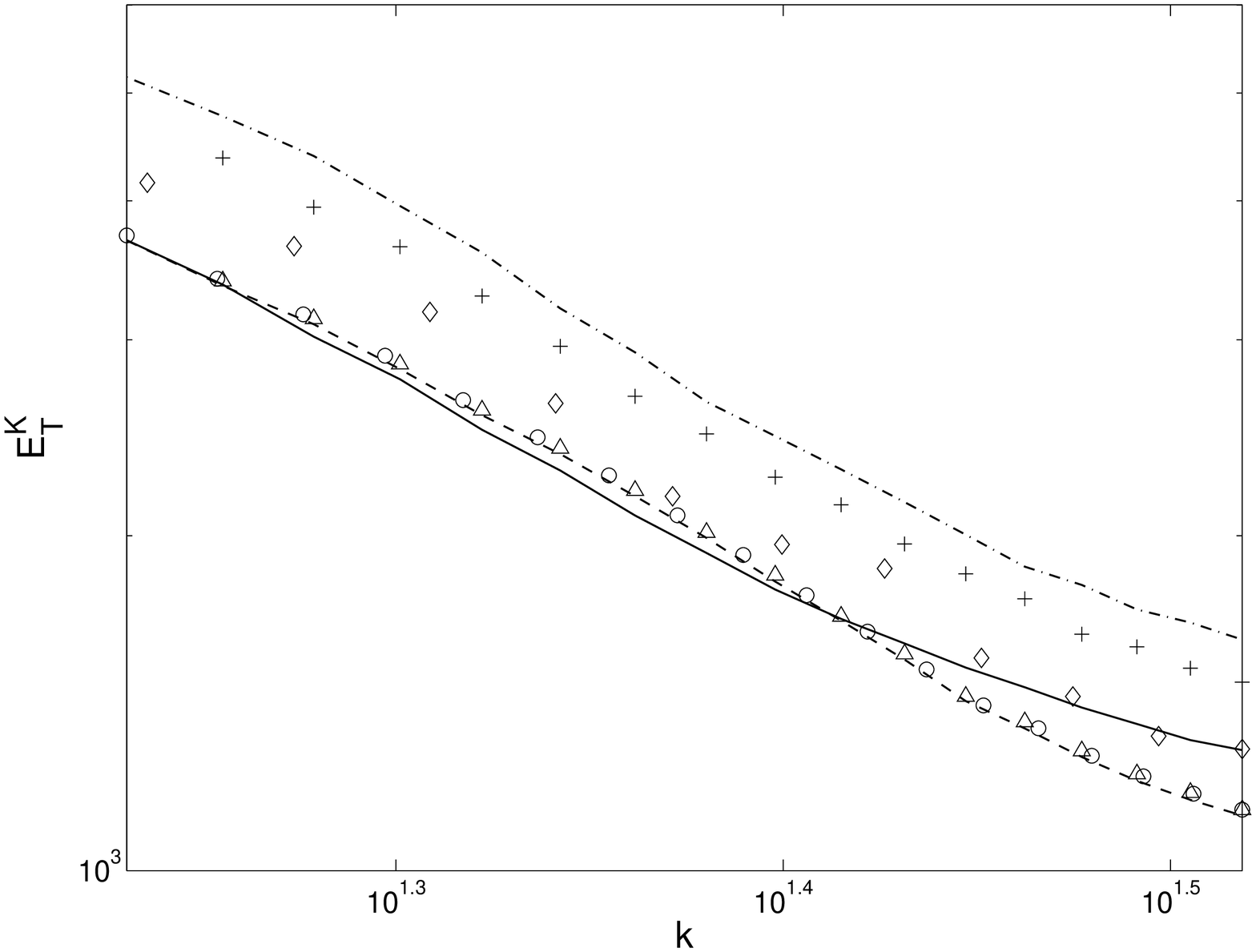}}
\caption {Total energy spectrum $E_{T}^{K}$ in enlargement zone of large values of wave number $k$ for various filter shapes. Symbols as in Fig.\ref{Fi:b_rms_filtr}}%
\label{Fi:tot_spect_filtr_big}
\end{figure}

The spectral distribution of the
kinetic and the magnetic energies that shows redistribution of
energy depending on wave number, i.e., at different scales. The investigation of inertial range properties is one of the main
tasks in studies of scale-similarity spectra of MHD turbulence.
Inertial range properties are defined as time averages over
periods of stationary turbulence conditions.
It is worth noting that the famous spectra of Iroshnikov-Kraichnan and Kolmogorov-Obukhov for MHD turbulence were obtained for the total energy. Total energy is the sum of kinetic and magnetic energy $E_{T} = E_{M} + E_{K}$. The spectra of total energy $E_{T}^{K}$ corresponding to these various cases are shown in Fig.\ref{Fi:tot_spect_filtr}. As expected from the theory of LES method, the main differences in the results are concentrated on the small (unresolved) scales. In order to observe these differences better, for clarity sake, Figure \ref{Fi:tot_spect_filtr_big} shows enlargement zone for large values of wave number $k$. It should be noted that the Gaussian filter is more sensitive to the parameter $\epsilon$ than the top-hat one for scale-similarity model in compressible MHD turbulence. From our calculations it can be seen that the 3-point filters give the worst results and the 5-point Gaussian filter demonstrates the best results (that is, best approximation to DNS) at $\epsilon=2$. However, the difference between these filters is still within $10\%$.

\section{Concluding remarks}
It appears that parameter $\epsilon$ which represents the ratio of the mesh size to the
cut-off lengthscale of the filter is an important parameter regarding the discrete filters of LES approach.
The present study summarized results concerning discrete filters for LES method of forced compressible MHD turbulent flows with the scale-similarity model. Scale-similarity parametrization has evident advantages in forcing compressible turbulence. Influences and effects of discrete filter shapes on the scale-similarity model were examined in physical space using a finite-difference numerical schemes. In this paper, the obtained results of numerical computations for LES were compared with the DNS results
of three-dimensional compressible forced MHD turbulent flows. The comparison between LES and DNS results was carried out regarding the time evolution of $b_{rms}$ and $u_{rms}$, and the total energy spectra of MHD turbulence. It was shown that the Gaussian filter is more sensitive to the parameter $\epsilon$ (the ratio of the mesh size to the cut-off lengthscale of the filter) than the top-hat filter for the scale-similarity model in compressible MHD turbulent fluid flow. Noteworthy result is that discrete filters produce more discrepancies for magnetic field. Therefore, it is important to choose correctly a filter using LES approach for modeling of forced compressible MHD turbulence. The 3-point filters at $\epsilon=2$ and $\epsilon=3$ give the least accurate results and the 5-point Gaussian filter demonstrates the best results at $\epsilon=2$. The difference between these filters is within $10\%$. As expected, the main differences in the results are concentrated on the small scales.

\begin{acknowledgment}

The work was supported by the
program P-22 of Russian Academy of Science Presidium "Basic problems in solar system studies".

\end{acknowledgment}

%

\end{document}